\documentclass[prl, notitlepage, superscriptaddress,  reprint, twocolumn]{revtex4-1}
\usepackage{enumerate}
\usepackage{amssymb}
\usepackage{amsmath}
\usepackage{graphicx}
\usepackage[usenames,dvipsnames]{color}
\usepackage[normalem]{ulem} 

\usepackage[colorlinks,bookmarks=false,citecolor=NavyBlue,linkcolor=OliveGreen,urlcolor=blue]{hyperref}
\newcommand{\be}{\begin{equation}}
\newcommand{\ee}{\end{equation}}
\newcommand{\ba}{\begin{aligned}}
\newcommand{\ea}{\end{aligned}}
\newcommand{\bw}{\begin{widetext}}
\newcommand{\ew}{\end{widetext}}

\newcommand{\bea}{\begin{eqnarray}}
\newcommand{\eea}{\end{eqnarray}}

\def\doi{http://dx.doi.org/}

\begin{document}
\title{Quantum information scrambling after a quantum quench}
\author{Vincenzo Alba}
\address{Institute  for  Theoretical  Physics, Universiteit van Amsterdam,
Science Park 904, Postbus 94485, 1098 XH Amsterdam,  The  Netherlands}

\author{Pasquale Calabrese}
\address{SISSA and INFN, via Bonomea 265, 34136 Trieste, Italy}
\affiliation{International Centre for Theoretical Physics (ICTP), Strada Costiera 11, 34151 Trieste, Italy}

\begin{abstract}

How quantum information is scrambled in the 
global degrees of freedom of non-equilibrium many-body 
systems is a key question to understand local thermalization. 
A consequence of scrambling is that in the scaling limit 
the mutual information 
information between two intervals vanishes at all times, i.e., 
it does not exhibit a peak at intermediate times.
Here we investigate the mutual information scrambling after a quantum quench 
in both integrable and non-integrable one dimensional 
systems. We study the mutual information between two intervals of finite 
length as a function of their distance. 
In integrable systems, the mutual information exhibits an algebraic 
decay with the distance between the intervals, signalling weak  scrambling. 
This behavior may be qualitatively understood 
within the quasiparticle picture for the entanglement spreading. 
In the scaling limit of large intervals, times, and distances between 
the intervals, 
with their ratios fixed, this predicts a decay exponent equal to $1/2$.
Away from the scaling limit, the power-law behavior persists, 
but with a larger (and model-dependent) exponent. 
For non-integrable models, a much faster decay 
is observed, which can be attributed to the finite life time of the quasiparticles:
unsurprisingly, non-integrable models are better scramblers.

\end{abstract}

\maketitle

\section{Introduction} 

The spreading of quantum information  is a central process for our 
understanding of non equilibrium many-body systems. 
A fundamental question, key to this work, is
how the quantum information encoded in the initial state
gets dispersed globally during the dynamics following a quantum quench.
A prominent idea,  originally 
formulated in the context of the information paradox in black 
holes, is that in generic quantum many-body systems the 
information about initial local correlations is mixed up during the 
dynamics, and after long times it cannot be recovered by performing 
local measurements, but {\it global} ones are 
required~\cite{hayden-2007,sekino-2008,shenker-2014,swingle} (scrambling scenario).  
Unfortunately, very few 
explicit results are available for realistic systems,  
although calculations in conformal field theories 
(CFTs) with large central charge~\cite{asplund-2014,bala-2011,asplund-2015,leich-2015}, 
holographic setups~\cite{allais-2012}, and mean-field-like models~\cite{syk}
provide useful insights. 
Several tools have been proposed to diagnose scrambling, 
such as the tripartite information~\cite{hosur-2016,oskar-2018,landsman-2018,pappa-18}, 
out-of-time-order 
correlators~\cite{shenker-2014,larkin-1969,maldacena-2016,bll-18,khemani-2018,sarang-2018}, and 
entanglement of operators~\cite{zanardi-2001,prosen-2007,prosen-2007a,znidaric-2008,pizorn-2009,dubail-2017,zhou-2018,jonay-2018,xu-2019,pal-2018,cdc-18,takayanagi-2018,nie-2018,bobenko,rgpp-19,hh-19}. 

Scrambling  is unanimously expected to have a different 
nature in integrable and generic, i.e., non-integrable, systems,
although a quantitative assessment has not been made and it is one 
of the main goals of this work.
To define scrambling it is useful to first consider intergrable systems. 
Integrable systems possess well-defined quasiparticles, which have 
a local-in-space nature, much alike to classical solitons, and infinite 
lifetime. After a quench from a low-entangled initial state 
these quasiparticles are created locally and uniformily in the 
system. As quasiparticles move ballistically, they spread the initial (EPR-like) 
correlations~\cite{calabrese-2005}. 
Note that initial correlations get ``dressed'' by non-trivial 
thermodynamic and many-body effects. For instance, 
the late-time properties of the quasiparticles after a quantum 
quench are described by an emergent thermodynamic macrostate 
(typically a Generalized Gibbs Ensemble (GGE)~\cite{rigol-2007,calabrese-2016,essler-2016,vidmar-2016,caux-2016},
in contrast to the thermal ensemble for non-integrable models \cite{D91,S94,R08,gogolin-2015,dalessio-2016,kauf}).
Hence, the entanglement entropy carried by the 
quasiparticles becomes the {\it thermodynamic} 
entropy of the stationary ensemble \cite{dls-13,collura-2014,nahum-17,AlCa17,nwfs-18,kauf}. 
In the appropriate space-time scaling limit, the entanglement 
evolution is quantitatively describable by a simple hydrodynamic 
framework (quasiparticle picture \cite{calabrese-2005}), for both
non-interacting and interacting systems~\cite{AlCa17,AlCal-a,AlCal-b,AlCal-c,AlCal-d,AlCal-e,fagotti-2008,mkz-17}.  

As we will clarified in the following,  
since the quasiparticles possess a non-trivial dispersion, as time progresses 
initial local correlations are spread (scrambled) 
over larger regions of the system. 
In order to better understand this scrambling let us consider two disconnected 
subsystems of finite length and at a certain distance. At a given time 
the correlation between the two systems, which can be 
measured by the mutual information, is proportional to the 
number of entangled pairs that at that time are shared between them. 
As the distance between the two intervals increases, the number of 
shared entangled quasiparticles (and, hence, the mutual information) 
decreases because the quasiparticles 
have different velocities (i.e., a non-trivial dispersion). This means 
that in order to reconstruct the initial local correlation 
one has to increase the size of the two intervals, i.e., 
considering more non-local observables. 
Note that local information is still 
somehow protected in integrable models because it is transmitted 
in a ``localized'' form via the quasiparticles. We anticipate that 
this implies that the mutual information exhibits a peak at 
intermediate times, which decays ``slowly'', i.e., as a power law, 
with the distance between the intervals.

Non-integrable systems are expected to be better scramblers: 
either they do not have stable quasiparticles or quasiparticles have finite 
lifetime. Moreover, in contrast with integrable models, scattering between 
quasiparticles is not elastic. 
This suggest that chaotic systems should loose memory of initial local 
correlations faster as compared to integrable ones. 
This should be reflected in the absence or in a fast decay of the mutual information 
peak at intermediate times, as suggested by holographic calculations. 
In this sense, the vanishing of the mutual information peak can be 
considered an operational definition of scrambling, although the physical 
reason behind scrambling is the absence of quasiparticles that could transport 
correlations in a local manner.

The question that we address here is: Is it possible to go beyond this qualitative scenario, 
characterizing quantitatively quantum information 
scrambling in integrable and non-integrable systems? 
Here we positively answer this question by
studying the mutual information between two distant 
intervals $A_1$ and $A_2$~\cite{amico-2008,calabrese-2009,laflorencie-2016} 
\begin{equation}
\label{mi}
I_{A_1:A_2}=S_{A_1}+S_{A_2}-S_{A_1\cup A_2}.
\end{equation}
(Here $S_X\equiv-\textrm{Tr}\rho_X\ln\rho_X$ is entanglement entropy of the subsystem $X$ in terms of its reduced density 
matrix $\rho_X$.)
We consider two intervals of equal length $\ell$ at a distance $d$. 
The mutual information always exhibits a well defined peak at intermediate times, 
but its features depend on whether the system is integrable or not,
reflecting the different degrees of scrambling. 
Indeed,  for large distance $d$ and at fixed $\ell$, 
the peak amplitude decays algebraically in $d$ in the integrable 
case and much faster (likely exponentially) for non-integrable systems. 
We mention that this is in agreement with recent holographic calculations, 
which suggest that in chaotic systems the mutual information peak at 
intermediate times is absent~\cite{asplund-2014,bala-2011,asplund-2015,leich-2015}.

%
\begin{figure}[t]
\includegraphics[width=1\linewidth]{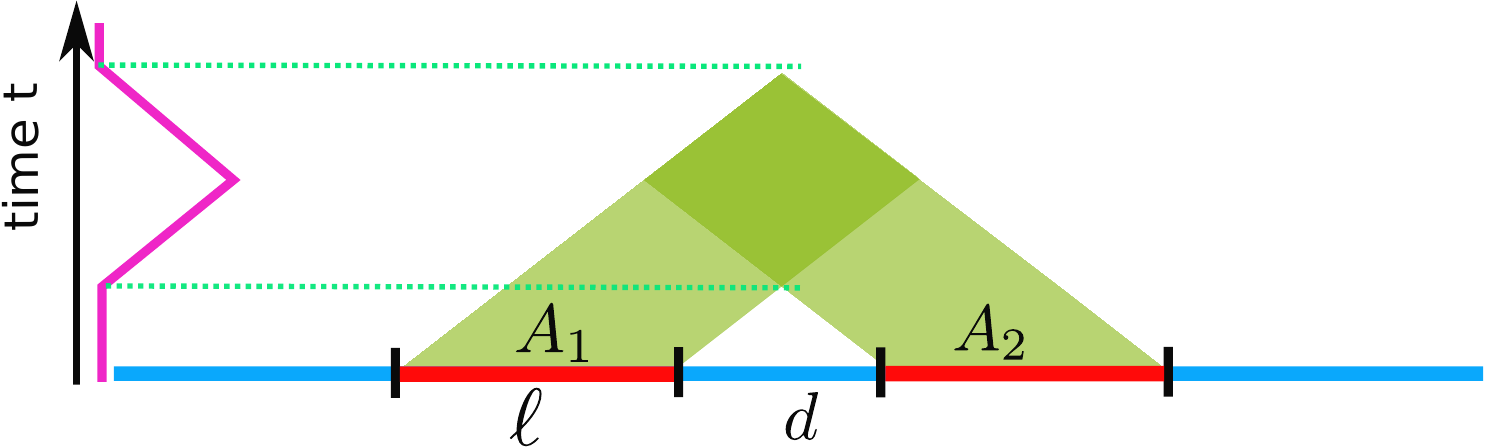}
\caption{ Dynamics of the mutual information $I_{A_1:A_2}$ between 
 two intervals in integrable models. 
 For quasiparticles with fixed velocity, at a given time $t$, $I_{A_1:A_2}$ is proportional to the 
 horizontal width of the intersection (darker area) 
 between the shaded regions at that time. 
 $I_{A_1:A_2}$ is shown on the vertical axis (purple line). 
 The generic result with a nontrivial dispersions $v_n(\lambda)$ in Eq. \eqref{quasi-full} follows from
 summing up over all possible quasiparticles. 
}
\label{fig:cartoon}
\end{figure}
%
\section{Scrambling and quasiparticles}
We start discussing the mutual information scrambling within the quasiparticle picture, cf. Fig.~\ref{fig:cartoon}. 
In generic integrable models there are, 
in principle, infinite species of quasiparticles~\cite{taka-book} labelled by an integer $n$. 
Quasiparticles of the same species are identified by a real parameter 
$\lambda$, called rapidity, that for non-interacting particles is the momentum. 
Within  each species, quasiparticles exhibit a non-trivial dispersion, i.e. a $\lambda$ dependent velocity $v(\lambda)$. 
We consider the situation in which 
only entangled pairs of quasiparticles with opposite rapidity $\lambda,-\lambda$ 
and opposite velocities $v_n(\lambda)=-v_n(-\lambda)$ are 
created after the quench (more complicated situations, for instance 
with entangled triplets, can be also treated~\cite{bertini-2018,bertini-2018a,bc-18}). 
According to the standard picture \cite{calabrese-2005},
the entanglement entropy is proportional to the number of quasiparticles shared between the subsystems of interest.

Let us first consider the case of a single quasiparticle with fixed velocity
$v(\lambda)=v,\,\forall\lambda$, as it happens in CFTs~\cite{calabrese-2005}. 
Given two intervals of equal length $\ell$ at distance 
$d$, the mutual information at time $t$  is the width of the shaded area in Fig. \ref{fig:cartoon}
 \cite{ctc-14,calabrese-2005}, i.e. 
$I_{A_1:A_2}=\textrm{max}(d/2,v t)+\textrm{max}((d+2\ell)/2,v t)-2\textrm{max}((d+\ell)/2,v t)$. 
$I_{A_1:A_2}$ is zero for $t<d/(2v)$, then it grows linearly in time up to $ t=(d+\ell)/(2v)$, after it decreases, still linearly, up to $t=(d+2\ell)/(2v)$ 
and for larger times it vanishes. 
Thus the maximum is attained for $ t=(d+\ell)/(2v)$,
the height of the peak is $\propto\ell$ and does not depend on $d$, signalling that scrambling is completely absent. 
Indeed, the absence of the mutual information peak 
has been considered a smoking gun for scrambling in CFTs with large 
central charge~\cite{asplund-2015} and holographic models~\cite{asplund-2014,bala-2011,allais-2012}. 

Let us know move to the realistic case of quasiparticles with a non-trivial dispersion. 
The mutual information is obtained by summing up the contributions of the 
different species of quasiparticles as~\cite{AlCa17,AlCal-a}. 
In the scaling limit, i.e., for any values of the ratios $d/\ell$ and 
$v_\mathrm{max}t/\ell$ (with $v_{\rm max}$ the maximum velocity of 
all quasiparticles) in the limit $d,\ell,t\to\infty$, the the 
mutual information is described by  
\begin{multline}
\label{quasi-full}
I_{A_1:A_2}=\sum_n I^{(n)}=\sum_n\int d\lambda s_n(\lambda)[\textrm{max}(d/2,v_n(\lambda) t)+\\
\textrm{max}((d+2\ell)/2,v_n(\lambda)t)-2\textrm{max}((d+\ell)/2,v_n(\lambda)t)], 
\end{multline}
Here $s_n(\lambda)$ is the quasiparticle contribution to the GGE entropy of the 
steady state ~\cite{AlCa17} and $v_n(\lambda)$  the quasiparticle velocity~\cite{fabian-v}. 
Both $s_n$ and $v_n$ can be calculated using thermodynamic Bethe ansatz~\cite{AlCal-a}.
According to Eq. \eqref{quasi-full},  $I_{A_1:A_2}$ is zero for $t<d/(2v_{\mathrm{max}})$, 
then it increases linearly up to $t\le (d+\ell)/(2v_{\mathrm{max}})$.
At later times it exhibits a short and slower increase until it reaches a maximum and finally it slowly decays at long times.
Both the growth after $t> (d+\ell)/(2v_{\mathrm{max}})$ and the asymptotic slow decay 
are due to the presence of slow quasiparticles. 
An intriguing feature of $I_{A_1:A_2}$ in integrable models is a multi-peak structure \cite{AlCal-a,mestyan-2017,pvcp-19}
as a consequence of the different maximum velocities of the diverse species.

In the scaling limit with $\ell\propto d$, it is clear that $I_{A_1:A_2}/\ell$ collapses on a scaling function of $t/\ell$ (or $t/d$); hence
the peak close to $t= (d+\ell)/(2v_{\mathrm{max}})$ does not depend on $d$ and there is  no actual sign of scrambling, 
as for the case with fixed velocity, as indeed tested in few instances \cite{AlCa17,ctc-14}. 

However, here we change the perspective and consider intervals of fixed 
length with increasing distance, i.e. $\ell\ll d$. 
In principle this limit is not captured by the quasiparticle picture 
because the quasiparticle picture is expected to hold when all length scales 
are large, including $\ell$. 
Formally, the result that we are going to derive holds in the limit 
$\ell,d\to\infty$ with fixed $\ell/d=\epsilon$ and $\epsilon$ is small. 

In fact, it is really illuminating to look at this regime 
within the quasiparticle picture.  Now, in 
Eq. \eqref{quasi-full} only the quasiparticles within a 
shell of width proportional to $\ell/d$ close to 
$v\approx v_{\textrm{max}}$ may contribute to the 
height of the peak of $I_{A_1:A_2}$. 
As a consequence, the peak of the mutual information 
vanishes as $d\to\infty$. Thus, even for integrable models 
there is a sort of weak scrambling, related to the non-trivial
dispersion of the quasiparticles. 

\begin{figure}
\includegraphics[width=0.99\linewidth]{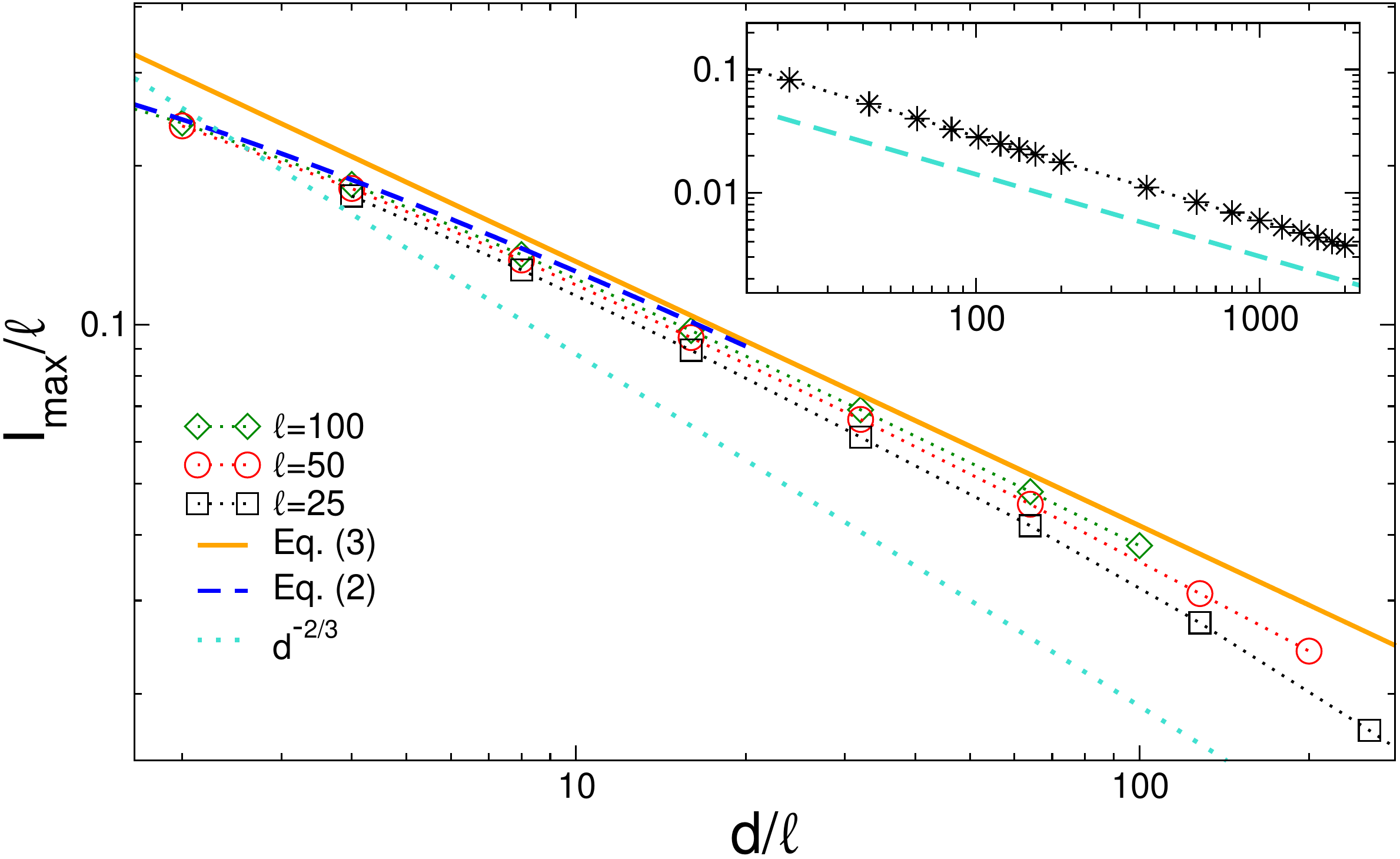}
\caption{ Mutual information after the 
 quench from the N\'eel state in the tight binding model: Rescaled 
 mutual information peak $I_\textrm{max}/\ell$
 as a function the  rescaled distance $d/\ell$ between the intervals. 
 The continuous  line is the quasiparticle prediction~\eqref{quasi-ld} 
 $\propto (d/\ell)^{-1/2}$ in the limit $d/\ell\to\infty$. 
 The dashed line $\propto d^{-2/3}$ is the result for finite $\ell$ and $d\to\infty$. 
 For any fixed $d/\ell$, upon increasing $d,\ell$ the quasiparticle 
 prediction Eq.~\eqref{quasi-full} (dashed line) is recovered, which, for large 
 $d/\ell$ becomes Eq.~\eqref{quasi-ld} (continuous line). For finite $\ell$ the 
 behavior as $d^{-2/3}$ is recovered in the limit $d\to\infty$ (dotted line). 
 Inset: two intervals with $\ell=5$ and $d\gg\ell$. 
 The dashed line is the behavior $d^{-2/3}$. 
}
\label{fig:peak}
\end{figure}

The above reasoning can be made quantitative by 
expanding up to second order $v_{n}(\lambda)$ and $s_n(\lambda)$ 
around the rapidity $\lambda_\textrm{max}$ of the fastest quasiparticle. 
We focus on the maximum $I^{(n)}_{\mathrm{max}}$ 
of the mutual information associated with quasiparticle species $n$. 
At the leading order in $\ell/d$, Eq.~\eqref{quasi-full} provides   
\begin{equation}
\label{quasi-ld}
I_{\textrm{max}}^{(n)}=\frac{4}{3}\left(\frac{2 v_{n,\textrm{max}}\ell}
{v_{n,\textrm{max}}'' d}\right)^\frac{1}{2}s_{n,\textrm{max}}
\ell,\quad d\gg\ell.
\end{equation}
In~\eqref{quasi-ld}, $s_{n,\textrm{max}}$ is the thermodynamic 
entropy of the fastest quasiparticle of the species $n$ with 
rapidity $\lambda_\mathrm{max}$, 
and  $v_{n,\mathrm{max}}''\equiv d^2v_n(\lambda)/d\lambda^2|_{\lambda_{\rm max}}$.  
We also assume that $v''_{n\textrm{max}}\ne 0$, that 
$v_{n}(\lambda)$ has only one maximum, and that 
$s_{n,\textrm{max}}\ne 0$.  
Hence, each quasiparticle species is responsible for a peak $I_{\textrm{max}}^{(n)}$ in the mutual information decaying for large distance as $d^{-1/2}$.
Two remarks are now in order. First, in the limit 
$d,\ell\to\infty$ with $\ell/d=\epsilon$ the size of the 
intervals increases upon increasing $d$, which leads to an 
enhancement of the mutual information between the two 
intervals compared to the situation with $d\to\infty$ 
at fixed $\ell$. This means that Eq.~\eqref{quasi-ld} 
has to be interpreted as an upper bound for the height 
of the mutual information peak between 
two intervals at fixed $\ell$ in the limit $d\to\infty$. 
This means that the exponent of the power-law 
decay of the mutual information peak can be larger than 
$1/2$. 
Indeed, we will see in a free-fermion model 
that for finite $\ell$ (i.e. of the order of 1), 
the mutual information decays as a power of $d$ for large $d$
with an exponent larger than $1/2$.
Second, it is important to discuss the regime of 
validity of Eq.~\eqref{quasi-ld}. Formally Eq.~\eqref{quasi-ld} holds in 
the limit $d,\ell\to\infty$ with  $\ell/d=\epsilon\to0$. 
Instead, for any finite but large $\ell$ there 
is a window with $1\ll\ell\ll d<\infty$ where Eq.~\eqref{quasi-ld} 
approximately describe the mutual information peak because the 
system is close to the scaling limit. For $\ell\to\infty$ 
Eq.~\eqref{quasi-ld} becomes exact in the limit 
$d\to\infty$ (see Fig.~\ref{fig:peak}).

%
%

\section{The model} To benchmark our results, we consider the spin-$1/2$ chain 
described by the Hamiltonian  
\begin{multline}
\label{xxz-ham}
H=\sum_{i=1}^L\Big[\frac{1}{2}(S_i^+S_{i+1}^-+
S_i^-S_{i+1}^+)+\Delta S_i^z S_{i+1}^z\Big]\\+
J_{nn}\sum_{\alpha=x,y,z}\sum_{i=1}^{L-2}S_{i}^\alpha S_{i+2}^\alpha+
h_x\sum_{i=1}^L S_i^x.
\end{multline}
Here $S_i^{+,-,x,z}$ are spin-$1/2$ operators. 
$J_{nn}$ is the strength of the next-nearest-neighbor interaction, $h_x$ a longitudinal magnetic field, and $\Delta$ an anisotropy parameter. 
For $J_{nn}=h_x=0$ the model is the XXZ chain, which is integrable by Bethe ansatz. 
For $J_{nn}=0$ and $h_x=1$, $H$ is integrable only at $\Delta=1$. 
For $J_{nn}\ne0$ the model is not integrable. For $J_{nn}=h_x=\Delta=0$, 
$H$ defines the XX chain, which is mappable onto the free-fermion tight binding
model $H=\sum_i (c^\dagger_i c_{i+1}+c_ic_{i+1}^\dagger)$, with 
$c_i$  standard fermionic operators. 
Here we only consider  the quench from the 
N\'eel state $|\textrm{N}\rangle\equiv \left|\uparrow\downarrow\uparrow\cdots\right
\rangle$. 

%
\begin{figure*}
\includegraphics[width=0.99\textwidth]{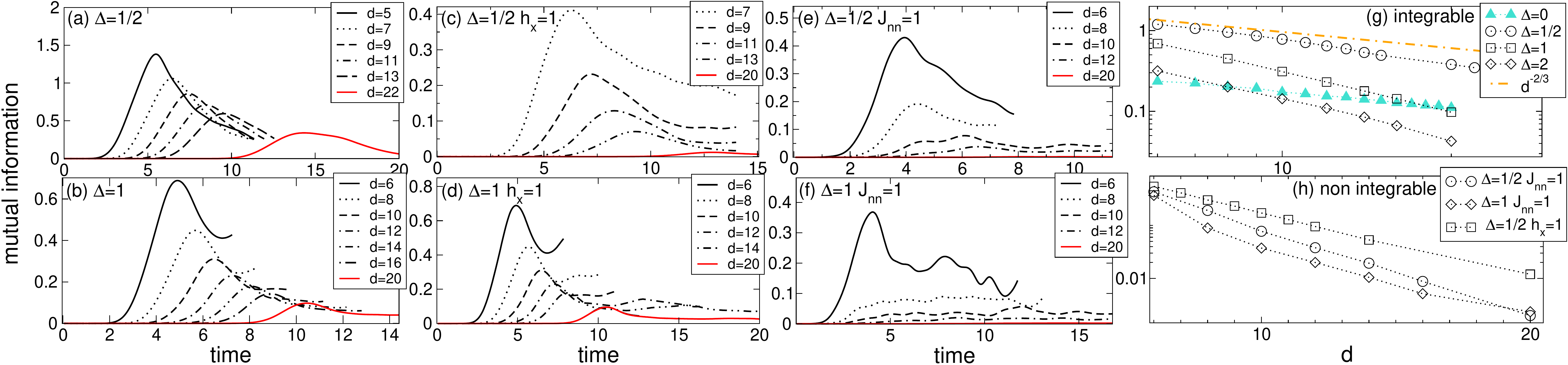}
\caption{Mutual information scrambling after the quench from 
 the N\'eel state: Integrable versus non integrable dynamics. 
 Mutual information $I_{A_1:A_2}$ between two 
 intervals of length $\ell=2$ at distance $d$, plotted as a function of time. 
 (a-b): Results for the XXZ chain with $\Delta=1/2,1$. 
 (c-d): The same as in (a-b) for the XXZ chain with a magnetic field $h_x=1$. 
 The model is non-integrable, except at $\Delta=1$  (d). 
 Notice in (c) the fast decay of the peak upon increasing $d$. 
 (e-f): Results for the XXZ with next-nearest-neighbor
 interaction  $J_{nn}=1$. Notice that already for $d\approx10$ 
 the peak is not visible.  
 (g-h): Height of the mutual information 
 peak as a function of $d$ for the integrable (g)  and 
 non-integrable (h) case. In (g), for all the quenches, $I_{A_1:A_2}$ 
 exhibits an algebraic decay $\propto d^{-\alpha}$. For the 
 XX chain $\alpha=2/3$ (dashed-dotted line) and for other 
 $\Delta$ we have $\alpha>2/3$. (h) The non-integrable case. 
 For all quenches, a fast decay is observed, compatible with an 
 exponential behavior. Note the logarithmic scale on the $y$-axis 
 only. 
}
\label{fig:panel}
\end{figure*}
%
\section{Free-fermion scramblers}
Let us first focus on the tight binding model. The fermionic mutual information dynamics is 
calculable from the time-dependent two-point correlation function
$C_{x,x'}$ restricted to $A_1\cup A_2$~\cite{pe-09}  (fermionic and spin mutual informations are different \cite{ip-09,fc-10}). 
In free-fermion language the N\'eel state is  
$|\textrm{N}\rangle=\prod_{j=1}^{L/2}c^\dagger_{2j}|0\rangle$. 
A straightforward application of Wick theorem yields
the time-dependent correlation function 
in the thermodynamic limit which reads 
\bea
\langle c^\dagger_x(t) c_{x'}(t)\rangle&=&\frac{\delta_{x,x'}}{2}+\frac{(-1)^{x'}}{2}
\int_{-\pi}^\pi\frac{dk}{2\pi}e^{i k (x-x')+4 i t \cos(k)} \nonumber
\\
\label{corr}
&=&\frac{\delta_{x,x'}}{2}+
\frac{(-1)^{x'}}{2i^{x-x'}}J_{x-x'}(4t), 
\eea
with $J_\alpha(x)$ the Bessel function of the first kind. 
Denoting with $\lambda_i$ the eigenvalues of the correlation matrix restricted to a subsystem $A$, the  
entanglement  entropy is $S_{A}=-\sum_{i} (\lambda_i\ln\lambda_i+(1-\lambda_i)\ln(1-\lambda_i))$~\cite{pe-09}, 
and the mutual information follows from the definition \eqref{mi}. 

It is instructive to consider first the mutual information between 
two fermions at distance $d$. The 2-by-2 correlation matrix $C_{x,x'}$ is just \eqref{corr}
with eigenvalues $\lambda_\pm$
\begin{equation}
\label{eig-1s}
\lambda_\pm=\frac{1}{2}\pm \frac{d^{-\frac{1}{3}}}{6^\frac{2}{3}\Gamma\big(\frac{2}{3}\big)}, 
\end{equation}
and so  $I_{A_1:A_2}=2^{2/3}/(3^{4/3}\Gamma^2(2/3))/d^{2/3}$: 
for free fermions, the peak of the mutual 
information for $\ell=1$ decays as $d^{-2/3}$. 
We are going to show that this behavior persists for larger, but finite, $\ell$.

For finite $\ell$, the mutual information peak is at
$t=(d+\ell)/4$, since $v_{\rm max}=2$. 
The asymptotic behavior of $I_{A_1:A_2}$ for $d\to\infty$ is obtained from $C_{x,x'}$.
The two points $x,x'$ are either in the same interval ($A_1$ or $A_2$) or in different ones. 
In the former case, the contribution of $k(x-x')$ in \eqref{corr} is negligible for finite $\ell$. 
Hence the integral \eqref{corr} for large $d$ is given by the stationary points at $k=0,\pm\pi$, i.e. 
\begin{equation}
\int_{-\pi}^\pi\frac{dk}{2\pi}e^{i d\cos(k)}\stackrel{d\to\infty}{\longrightarrow}\frac{\sqrt{2}}{\sqrt{\pi d}}
\cos\big(\frac{\pi}{4}-d\big).
\end{equation}
If $x$ and $x'$ are in different intervals $A_i$, then 
$x-x'\propto d$, and the integral is dominated by the stationary point at 
$k=-\pi/2$. 
Crucially, the saddle point contribution vanishes at $k=\pi/2$,  
and one has to consider the order $k^3$, which gives 
\begin{equation}
\label{asy}
\int_{-\pi}^\pi \frac{dk}{2\pi}e^{i d k+i d\cos(k)}\stackrel{d\to\infty}{\longrightarrow}
\frac{1}{d^\frac{1}{3}}\int_{-\infty}^\infty \frac{e^{\frac{1}{6} i k^3}}{2\pi}
=\frac{-6^\frac{1}{3}}{d^\frac{1}{3}\Gamma\big(-\frac{1}{3}\big)}. 
\end{equation}
The exponent $1/3$ appearing in~\eqref{asy} is ubiquitous in free-fermion 
models and it is related to the Airy processes \cite{eisler-2013,clm-15,ms-16,vsdh-16,dlms-18,lms-18,stephan-2019}. 
At this point, the mutual information is dominated by the elements of $C_{x,x'}$ coming from $x,x'$ in different intervals and,
since $\ell$ is finite, the same  $d^{-2/3}$ behavior holds in general.
This is explicitly tested in the inset of Fig.~\ref{fig:peak} for $\ell=5$ where the decay $d^{-2/3}$ is observed neatly.

How is the quasiparticle prediction~\eqref{quasi-ld} recovered for larger $\ell$?
This question is investigated  in Fig.~\ref{fig:peak} by diagonalizing the correlation matrix numerically.
We report the height of the mutual information peak after the N\'eel quench for different $\ell$. 
It is evident that the curves for $\ell=25,50, 100$ show a crossover from the $d^{-1/2}$ scaling 
for not so large $d$, to a truly asymptotic $d^{-2/3}$ decay for very large $d$.
The quasiparticle prediction~\eqref{quasi-ld} well describes the data in a
preasymptotic regime, as expected.

\section{Interacting integrable and non-integrable scramblers}
We finally consider the  effect of interactions and of integrability-breaking perturbations.
The quench from the N\'eel state with several parameters of the Hamiltonian \eqref{xxz-ham} 
is studied by means of  tDMRG~\cite{uli-2011} simulations.
It is however computationally very demanding to calculate the 
entanglement  entropy of two disjoint intervals.
The computational cost is $\propto\chi^3$, with $\chi$ the 
bond dimension which must grow exponentially with $\ell$. 
Here we use $\chi\lesssim 1500$. 
Thus, we restrict ourselves to simulate small intervals with $\ell=2$ placed at the ends of an open chain
and so we can only explore the regime $ \ell\sim O(1)$, without accessing the possible quasiparticle regime.
All our data are reported in Fig.~\ref{fig:panel}.    
In panels (a), (b), and (d) we focus on the integrable case. 
Quite generically, $I_{A_1:A_2}$ exhibits a clear peak at intermediate times. 
The peak decreases as a function of $d$ 
and it remains visible even at large $d$. 
(Interestingly, at $\Delta=1/2$ for $d=22$ a second peak 
appears, reflecting the presence of two species of quasiparticles. 
However, $d=22$ is not large enough to resolve the two peaks neatly.)

The picture changes dramatically upon breaking integrability. 
In Fig.~\ref{fig:panel} (c) we break integrability by switching on a 
longitudinal magnetic field $h_x=1$ (cf.~\eqref{xxz-ham}). 
Now a peak is visible only for small $d$, whereas 
it decays rapidly at large $d$. This suggests a much faster decay of the 
mutual information in non-integrable models. 
We now move to a different integrability-breaking 
perturbation by setting $J_{nn}=1$ in~\eqref{xxz-ham}. 
Fig.~\ref{fig:panel} (e-f) shows results for $J_{nn}=1$ and $\Delta=1/2,1$. Surprisingly, a peak is 
only visible for $d=6,8$, and it rapidly melts into a broad plateaux  at 
larger $d$. The height of the plateaux decreases quickly with $d$. 

In Fig.~\ref{fig:panel} (g) and (h) we analyze the decay of the mutual information peak/plateaux 
as a function of $d$. 
The integrable cases are summarized in (g). 
For $\Delta=0$, i.e., for the XX chain, the expected decay $d^{-2/3}$ is  visible already for relatively small $d$. 
For other values  of $\Delta$ a clear power-law behavior is found. For instance, at $\Delta=1/2$ the data 
follow a $1/d$ behavior, while for $\Delta=1,2$ they suggest a faster algebraic decay. 
However, larger values of $d$ would be needed to extract the correct power-law reliably for all $\Delta$.
For the non-integrable case in Fig.~\ref{fig:panel} (h)  a faster decay is 
observed as compared with the integrable case in Fig.~\ref{fig:panel} (g). 
The data are compatible with an exponential decay, in spite of the fact that we have only one decade of data for $d$.
Thus our data are compatible with a strong-scrambling scenario for non-integrable models. 
Note that in the kicked Ising chain, which is regarded as a maximally chaotic model, 
the mutual information peak is completely absent~\cite{bertini-2019}.
In chaotic random circuits, the mutual information is exponentially small for $\ell<d$ \cite{nahum-17}.

\section{Conclusions}
We investigated the information scrambling after a quantum quench in integrable and non-integrable 
systems, focusing on the mutual information between far apart intervals of fixed lengths. 
While the standard configuration in the quasiparticle picture with $\ell\propto d$ has an 
enormous computational cost, intervals of fixed lengths can be simulated easily, as noticed long ago \cite{lk-08}.
We found that for integrable systems the mutual information decays as a power-law of the 
distance between the intervals.
For non-integrable models a faster decay, compatible with an exponential, is  observed.

Our work motivates further studies in many new directions. 
First, it is important to corroborate the correctness of our 
conclusions for other systems and different quenches, both analytically and numerically.
Then, it is natural to wonder whether the model dependent exponent for the decay of the 
mutual information  in integrable systems can be obtained analytically. 
An intriguing possibility would be to understand whether this exponent 
can distinguish between interacting and free integrable systems \cite{s-17}. 
Our results may suggest that in interacting theories the exponent is larger than in free ones.
Concomitantly, further checks of the $d^{-2/3}$ decay in other non-interacting 
systems are necessary to assess its universality. 
A natural question is also whether a crossover between algebraic and exponential decay can take 
place in models with unstable but long lived quasiparticles, as for instance in prethernalization scenario \cite{EsslerPRB14,fc-15,BEGR:PRL,BEGR:long}
or in confining models \cite{kctc-17,mrw-17, jkr-18,llt-19,smg-19}. 
Finally, it would be interesting to investigate 
the mutual information scrambling in those models 
without a maximum quasiparticle velocity, such as 
integrable non-relativistic quantum field theories.

\paragraph{Acknowledgements.}
PC acknowledges support from ERC under Consolidator grant number 771536 (NEMO).
VA acknowledges support from the D-ITP consortium, a program of the NWO.
Part of this work has been carried out during the workshop  ``Entanglement in quantum systems'' at the Galileo Galilei Institute (GGI) in Florence.

\end{document}